\documentclass[a4paper]{jpconf}
\usepackage{graphicx}
\usepackage{url}

\def\Msun{\mbox{$M_\odot$}}
\def\ssec#1{\subsection{#1}}
\def\etal{{\it et al.}}

\begin{document}
\title{Astronomy and astrophysics with gravitational waves in the Advanced Detector Era}

\author{Alan J.~Weinstein \\
for the LIGO Scientific Collaboration and the Virgo Collaboration}

\address{California Institute of Technology, Pasadena, California 91125, USA}

\ead{ajw@ligo.caltech.edu}

\begin{abstract}
With the advanced gravitational wave detectors coming on line in the
next 5 years, we expect to make the first detections of gravitational
waves from astrophysical sources, and study the properties of the
waves themselves as tests of General Relativity. In addition, these
gravitational waves will be powerful tools for the study of their
astrophysical sources and source populations. They carry information
that is quite complementary to what can be learned from
electromagnetic or neutrino observations, probing the central
gravitational engines that power the electromagnetic
emissions at the outer layers of the source. 
Preparations are being made to enable near-simultaneous
observations of both gravitational wave and electromagnetic
observations of transient sources, using low-latency search pipelines
and rapid sky localization. We will review the many opportunities for
multi-messenger astronomy and astrophysics with gravitational waves
enabled by the advanced detectors, and the preparations that are being
made to quickly and fully exploit them.  
\end{abstract}

\section{Introduction}

In the coming years, advanced ground-based gravitational wave (GW) detectors 
will be coming online, observing the sky as a detection and analysis network.
We refer to this as the Advanced Detector Era (ADE).
The first detections, followed by a growing catalog of sources,
will be a trove of astrophysical information about the properties
of these extremely energetic systems.
In this talk and paper, I briefly review some of the many ways that
gravitational waves can inform us about those properties.

There are many promising sources for gravitational waves.
Transient sources include 
binary mergers, core-collapse supernovae, magnetar flares, 
and pulsar glitches.
Some of these are believed to be the progenitors of gamma ray bursts (GRBs)
or soft gamma repeaters (SGRs).
Long-lived sources include continuous waves from rapidly spinning
neutron stars,
and the stochastic background from the early universe. 
The scope of this discussion will be limited mainly to binary mergers,
with brief mention of other transient sources;
we will refer the reader to excellent talks on many of these other
subjects by speakers at this conference.

\section{The GW detector network in the advanced detector era}

\ssec{The initial detector era:}
The initial LIGO~\cite{iLIGO} and Virgo~\cite{iVirgo} 
detectors completed their science runs in 2007.
The ``enhanced'' LIGO~\cite{eLIGO} and Virgo~\cite{iVirgo} 
detectors ended data taking in October 2010. 
As of summer 2011,
the H2 detector at LIGO Hanford Observatory (LHO) and 
the L1 detector at LIGO Livingston Observatory (LLO) are dismantled
in preparation for Advanced LIGO installation.
The H1 Enhanced LIGO detector at LHO is conducting a squeezed vacuum experiment in Fall 2011, 
after which, it too will be dismantled.
During summer 2011, the Virgo~\cite{iVirgo} 
and GEO-HF~\cite{GEOHF} detectors are taking science data;
see the talk at this conference by Mirko Prijatelj on GEO-HF.
The Virgo detector will end observations in preparation for 
Advanced Virgo by the end of 2011.

\ssec{Astrophysics with the initial detectors:}
The years 2004 through 2011 saw the publication of 65 papers~\cite{LSCpapers}
documenting the scientific results obtained from the 
initial and enhanced LIGO, GEO600 and Virgo detectors.
These also include joint publications with the TAMA 
Collaboration in Japan (who operated a 300 meter interferometric
detector near Tokyo~\cite{TAMA}) and the
AURIGA and ALLEGRO bar detector collaborations~\cite{BARS}.
These papers reported on the performance of the detectors,
and the results of searches (all negative) for gravitational waves
from compact binary inspiral, merger and ringdown; 
GW bursts from supernovae, massive mergers, or magnetar excitations;
continuous GWs from known pulsars or spinning neutron stars;
and stochastic GWs from the early universe or unresolved astrophysical sources.
We will discuss the prospects for detecting and studying
GWs from transient sources (binary mergers and bursts) 
with the advanced detector network in Section 3 below.

\ssec{The advanced detector era:}
The construction of the Advanced LIGO (aLIGO,~\cite{aLIGO}) detectors
is around 50\%\ complete, and installation has begun at LHO and LLO.
Advanced LIGO construction will complete in 2014. 
Commissioning will commence, and first science runs are expected to commence
in 2014--2015 at Hanford and Livingston.
The Advanced Virgo detector~\cite{aVirgo}  in Cascina, Italy
and LCGT~\cite{LCGT} in Japan will also come online in the same time-scale;
see the talk at this conference by Giovanni Losurdo.
It should be kept in mind that it took several years for the initial LIGO detectors
to achieve design sensitivity. The advanced detectors are far more complex,
and although the commissioning team is far more experienced, it still may take
substantial time for the advanced detectors to reach their design sensitivity.
Note also that work is already in progress to design third-generation detectors
which could improve sensitivity with respect to the coming advanced detectors,
by as much as another factor of 10; see the talks at this conference by 
Stefan Hild and B.~Sathyaprakash.

\ssec{The third aLIGO detector:}
LIGO Laboratory is constructing three 4 km advanced detectors,
two for installation at LHO and one at LLO.
However, as discussed below, a far more powerful global detector network 
is possible if the third aLIGO detector is installed not in LHO
but in a distant observatory site such as western Australia~\cite{LIGO-A}:
see the talk at this conference by David McClelland; or India~\cite{LIGO-I}:
see the talk at this conference by Tarun Souradeep.
(Note that as of November 2011, the Australia option is no longer feasible
due to lack of funding commitment from the Australian government).

\ssec{The aLIGO strain noise spectrum:}
The advanced interferometric detectors are largely quantum noise limited over much
of the detection band. They can be operated at high laser power (up to 125 watts)
to reduce the shot noise at high frequencies (above 100 Hz) at the expense
of increased radiation pressure noise below around 50 Hz; or vice versa at lower power.
Furthermore, the advanced detectors feature a ``signal cavity'' at the asymmetric port
of the interferometer (where the GW signal is extracted). When that cavity
is tuned to the main laser carrier frequency, the detector operates in 
``broadband'' mode. When detuned, it can improve sensitivity over some frequency band
at the expense of others, so it can be optimized for binary neutron star signals
or signals from higher mass binaries.
Example anticipated aLIGO and aVirgo sensitivity curves can be found in 
Ref.~\cite{aLIGOcurves,aVirgocurves}.

\ssec{The global network:}
The global advanced detector network includes the two LIGO detector sites,
GEO-HF, Virgo, LCGT, and possibly LIGO-India. A detector network
enables more confident detection by requiring the same or similar signal 
in widely separated detectors. More importantly, a network is required in order
to locate the sources in the sky, since individual detectors have essentially 
no ability to determine the source direction for transient signals
(long-direction continuous wave sources can be well wocalized with only
one detector due to the Doppler shift associated with the detector motion
relative to the source). Individual detectors are 
laser interferometers with quadrupole antenna patterns; a detector network
acts as a gravitational-wave interferometer, determining source direction
through timing and amplitude information of a wave detected coherently 
across a global baseline.

\ssec{The need for a detector in Australia or India:}
The LIGO and Virgo detectors define a plane,
and sky localization resolution is good for sources lying outside that plane,
poor for sources near that plane. The addition of LCGT in Japan 
helps greatly, but that detector is also close to the same plane,
now forming a larger triangle. 
The addition of a detector in Australia or India forms a 
global tetrahedron, with much improved resolution in all directions;
see the talk at this conference by Linqing Wen.
To first order, LCGT improves East-West localization of the source, while
LIGO-Australia or LIGO-India improves North-South localization.
By coherently analyzing the data from all detectors as a single global network,
one can improve sky localization accuracy to the level of a few square degrees
over almost all of the sky~\cite{Bose,Nissanke,Wen,Fairhurst,Aylott,VitaleML},
for transient signals with SNRs well above the detection threshold.
In addition, a distributed detector network with non-aligned 
antenna response patterns allows one to
disentangle the two polarizations of the waves,
from which one can measure the orientation of the source
with respect to the detectors and thus the intrinsic amplitude of the 
waves at the source. This is crucial information for understanding
the astrophysical properties of the source:
the luminosity distance for standard sirens like binary systems;
the GW energy release for core-collapse supernovae; or
the non-sphericity of neutron stars emitting continuous waves.
The opportunity to relocate
one LIGO detector in Australia or India will greatly enhance the 
scientific output of GW observations, 
especially for transient signals;
please see the references above for analysis-dependent 
quantitative results on the improvement expected in the sky localization. 

\section{Astronomy and astrophysics with compact binary mergers}

Good source localization allows one to look for other forms of radiation
coming from that location, nearly simultaneously with the GWs:
optical, radio, X-ray or gamma electromagnetic radiation, or neutrinos.
Such multi-messenger signals may be expected from sources containing 
ordinary atomic or nuclear matter.

\ssec{The dark side:}
However, GWs carry information about strong, rapidly changing gravitational fields, 
even if there is no ordinary matter present to emit other forms of radiation --
what Kip Thorne calls ``the dark side of the universe''.
A binary black hole inspiral, merger and ringdown may not be accompanied 
by any other form of detectable radiation.
An isolated, rapidly spinning neutron star, even nearby, 
may not be observable in EM radiation.
But the GW waveform itself carries a wealth of information about the system, 
much more than what can usually be obtained from EM observations alone.

\ssec{Binary parameters:}
For a binary merger, this includes: the masses of the component objects;
the orientation of the binary orbit with respect to the detectors (two angles);
the sky coordinates of the system (two angles);
the luminosity distance; and the time and phase of merger (as observed on the Earth) --
a total of nine parameters.
In general, the compact components of the binary (neutron stars
and/or black holes) are spinning, and information about the magnitude 
and orientation of those two spin vectors is encoded in the waveform.
For example, if the component objects have high spin, aligned with the
orbital angular momentum, it will take more time and more gravitational 
wave cycles for the system to radiate away enough angular momentum
in order to merge (``hang-up'' and ``spin-flips''). 
The six components of the component spins bring the number of parameters
up to 15. If the orbit is eccentric, there are three more parameters;
but by the time a binary orbit enters the frequency band of ground-based detectors,
the extra radiation as the orbit passes through periastron
effectively circularizes it. Tidal distruption of neutron stars (see below)
adds more (model-dependent) parameters.

It is rare for radiation from an astronomical object
to carry so much information about the source. 
Electromagnetic radiation is typically generated from the 
incoherent motion of electrons in thermal or explosive events;
gravitational radiation comes from the coherent motion
of large (in LIGO's case, stellar-mass) objects.

\ssec{Waveform models:}
Extracting this information requires accurate models of the source.
For binaries, this is the relativistic gravitating two-body problem.
In the early phases of the inspiral, the waveform is well understood 
in the context of the post-Newtonian expansion~\cite{PPN}.
For the late stages of inspiral and merger, recent spectacular advances
in numerical relativity~\cite{CBCNR} have provided waveforms 
for a range of binary parameters, and semi-analytical and phenomenological
models have allowed us to interpolate (and even extrapolate) 
to cover the space of binary parameters;
see talks at this conference by Frank Ohme and Michael Boyle.

\ssec{Binary mass range:}
The parameter space is large: 9-dimensional if spin is neglected,
or 15-dimensional if it is included, as it needs to be.
Component masses can range from 1 \Msun, through the typical neutron star mass
range around 1.35 \Msun, and higher (where \Msun\ is the solar mass).
Compact objects with mass above
3 \Msun\ are expected to be black holes. Astrophysical black holes
that form from core collapse of very massive stars are expected to be as massive
as 20 \Msun\ or more. Supermassive black holes with masses of
$10^6$ \Msun\ are also known to exist; and it may be surmised that 
intermediate mass black holes with masses of 20 -- 1000 \Msun\ might 
also exist, perhaps formed via mergers of smaller black holes. 
Any and all of these may form binaries that may merge via gravitational radiation
within the age of the universe.

\ssec{Frequency at merger:}
Binaries with total mass M (in solar mass units) reach their
innermost stable circular orbit before merger when the GW frequency reaches
around 4000 Hz/M. Low mass systems (with total mass below around 5 \Msun) 
merge at high frequencies outside of the most sensitive band
of the advanced detectors (roughly 20 -- 1000 Hz),
while the merger of higher mass systems is in the most sensitive band.
Above 100 \Msun, only the last part of the waveform from the ringdown
of the final merged black hole is in band.
In the low mass case, post-Newtonian templates are sufficient to capture the signal;
in the high mass case, damped sinusoids are suitable templates.
In the intermediate case (total mass of 5 -- 100 \Msun),
models~\cite{EOBNR,Phenom}
that match the inspiral to the ringdown with a merger phase
tuned to numerical relativity waveforms are needed. 
These latter waveforms do not yet include the effects of spin.

\ssec{Sensitivity and rates:}
Using these models as templates for searches, one can estimate
the sensitivity of the advanced detectors to binary mergers
as a function of distance and total mass.
Experience from the initial LIGO-Virgo observations suggests that 
reliable detection (with false alarm probability of less than around
$10^{-5}$ for a first discovery)
is possible with an average SNR per detector of around 8.
Averaging over binary sky location and orbital orientation, 
the aLIGO detector can register a signal with SNR of 8 or more 
from a binary neutron star merger to a distance of around 200 Mpc,
and from a binary black hole system (with component masses of 10 \Msun\
each) to a distance of almost 1000 Mpc~\cite{rates}.
For all masses of interest, the sensitive distance is sufficient
to expect that the density of sources (galaxy mass or luminosity) 
per unit volume is uniform and constant (roughly 0.01 Milky Way
Equivalent Galaxies per Mpc$^3$)~\cite{rates, galaxy}. Using population
synthesis models anchored to the handful of observed binary neutron stars
in the Milky Way, this leads to predicted detection rates 
in the advanced detector era of around 40 binary neutron star mergers
and 20 binary black hole mergers per year (with uncertainties of a factor 10 or
more in both directions)~\cite{rates}.

\ssec{Parameter estimation:}
Once binary merger events are observed, it is possible to extract their
parameters;
see the talks at this conference by John Veitch and Vivien Raymond.
In practice, the signals will be buried in noisy detector data,
so the data from all detectors in the network are filtered through 
a coherent set of model waveforms (``templates''), which depend in known ways
on the parameters enumerated above, and a network SNR is computed. 
One explores the parameter space to maximize the SNR with respect to the parameters.
But because the parameter space is so large,
some efficient algorithm is employed, such as Markov Chain Monte Carlo~\cite{MCMC}
(MCMC) or nested sampling~\cite{Nested}. Even then, these calculations
are computationally demanding;
Parameter estimation on hardware injections in the 2009--2010 LIGO-Virgo observations
took many days per event on clusters of computers.
At present, run times have been reduced to hours for non-spinning signals 
and more work is in progress to speed up these computations dramatically.

\ssec{Parameter accuracy:}
The waveform carries information about both component masses,
but the phase evolution of the waveform depends to first order only on the combination
known as the chirp mass ${\cal M} \equiv M_{tot} \eta^{3/5}$,
where $M_{tot} = m_1+m_2$ is the total mass and
$\eta = m_1 m_2 / M_{tot}^2$ is the dimensionless symmetric mass ratio. 
The chirp mass is well measured (for example,
a system with a chirp mass of 5 \Msun\ at 30 Mpc can be determined 
by the initial detector network with an
accuracy of $\sigma_M \simeq 0.1$ \Msun, 
see the talk  at this conference by Veitch)
while $\eta$ is measured relatively poorly (for this example,
with an accuracy of $\sigma_\eta \simeq 0.02$),
leading to an extended measurement confidence band in $m_1$, $m_2$ parameter space.
Experience from hardware injections in the 2009--2010 LIGO-Virgo observations
inform us that the mass accuracy depends strongly on the waveform
accuracy (eg, one gets different results from 2.5-order and 3.5-order post-Newtonian
waveforms)~\cite{S6PE}. The dependence on spin is weaker.

\ssec{Sky position:}
As discussed above, the ability to measure sky position of the source 
depends greatly on the number and location of detectors in the network,
as well as on the type of waveform (frequency content) and, of course, the SNR.
Burst events near detection threshold in the initial detector network
can have 90\%\ confidence level
localization area of more than 100 square degrees, typically in
multiple disjoint regions and arcs across the sky~\cite{lumin}.
Even in the best of circumstances (high SNR for a 1 kHz signal in 
many advanced detectors),
sky localization is diffraction limited at not much better than 
a square degree~\cite{Bose,Nissanke,Wen,Fairhurst,Aylott,VitaleML}.

\ssec{Luminosity distance:}
Gravitational waves from binary inspiral are ``standard sirens''~\cite{siren},
making them useful tools for measuring the Hubble constant in the local
universe (redshift $z\approx 0$) with very different (and potentially small)
systematic errors with respect to measurements using standard candles
like Type 1a supernovae;
see talk at this conference by Walter Del Pozzo.
Their amplitude at the detector depends on the luminosity distance to the source,
but also on the sky position and orbit orientation;
there is a strong degeneracy between the luminosity distance and the
inclination $\iota$ of the orbit with respect to our line-of-sight.
The orbit orientation can be measured using a network of detectors
with different sensitivities to the two GW polarizations (plus $+$ and cross $\times$),
since these depend differently on $\cos\iota$.
In the best of circumstances (high SNR in many non-aligned detectors),
achieving 1\%\ resolution on luminosity distance
requires the detectors' amplitude calibration uncertainty to be of the 
same order~\cite{siren,Vitale};
see the talk at this conference by Salvatore Vitale.
In Enhanced LIGO, the detectors' calibration uncertainty was more like 10\%,
but improvements are expected for aLIGO.

\ssec{Component Spins:}
Waveforms from binaries with massive spinning components 
will be modulated in phase and amplitude.
The orbital angular momentum $L$ will precess around the total $J=L+S$,  
modulating the orientation of the orbital plane;
and thus the amplitude at the detector. 
The frequency evolution will also be modified (see discussion of orbital hang-up, above).
By making use of a sufficiently accurate waveform model,
one can estimate the two 3-dimensional spin vectors
(relative to the total or orbital angular momentum)
from the observed waveforms.
Much effort has been going into the development 
of the waveform models that incorporate spin~\cite{spinWF}, 
as well as the parameter estimation codes that explore spin parameter space
with those waveforms~\cite{MCMC};
see talks at this conference by Riccardo Sturani, Emma Robinson, Bernard Kelly,
P. Ajith, and Vivien Raymond.

\ssec{Tidal disruption of neutron stars in merging binaries:}
The GW waveforms from compact binaries deviate from the point-particle limit
as the objects approach merger. For binary black holes, this is in principle
computable in numerical relativity from the equations of General Relativity.
If one or both of the compact objects are neutron stars, they will become
tidally disrupted near merger, severely distorting the waveform 
at high frequencies (for binary neutron stars, above around 500 Hz).
This is a complicated process, involving not just strong field highly-dynamical gravity,
but also nuclear physics, magnetic fields,
and the formation of an accretion disk.
The first simulations are now in hand~\cite{tidalsim};
see the talk at this conference by Kenta Kiuchi.
To lowest order, however, the frequency at which the distortion sets in
is governed by the size (compactness) of the neutron star.
This in turn depends on the 
microscopic laws governing the equation of state 
(relation between pressure and density, or compressibility)
of nuclear matter in the stars; properties of great interest to physicists interested in
nuclear matter at the extreme conditions of neutron stars.
``Soft'' equations of state result in more compressible matter and thus more compact
stars, which disrupt at tighter orbits (higher frequencies);
``hard'' equations of state are the opposite.
Since advanced detectors have noise curves that get worse at frequencies 
of 500 Hz and above, much effort is going into improving detector
sensitivity at higher frequencies, and in measuring the
neutron star compactness at the lowest frequencies that can 
carry that information~\cite{tidalanal};
See talks at this conference by Jocelyn Read, Filippo Galeazzi, and Benjamin Lackey.

\ssec{Testing General Relativity and the no-hair theorem:}
The waveform from the inspiral phase of binary merger 
is well predicted in post-Newtonian expansion
of General Relativity (GR), so observations can test that prediction;
any deviations can be interpreted as a breakdown of GR that point to a 
deeper theory. Those deviations can be parameterized in the context
of a specific model, or in a model-independent 
(phenomenological) way~\cite{Iyer,TLi};
see talks at this conference by Bala Iyer and Tjonnie Li.
The merger into a single but highly perturbed black hole
will cause it to ring down in a set of quasi-normal modes~\cite{QNM}.
The mode frequencies and damping times are all predictable in 
black hole perturbation theory.
If multiple modes can be observed, their consistency with each other 
according to the predictions is a powerful test of the black hole
no hair theorem~\cite{nohair};
see talk at this conference by Ioannis Kamaretsos.

\ssec{Beyond GR:}
Many tests of GR have been performed in the weak field regime,
and quantified in terms of the parameterization developed
by Clifford Will~\cite{Will}.
The predictions of GR involving gravitational waves have yet
to be confirmed or tested (except indirectly using the binary pulsars).
Techniques are being developed to directly measure the
speed of gravitational waves (equal to the speed of light according to GR).
In quantum language, this constrains or 
measures the mass of the graviton~\cite{graviton}.
Similarly, GR predicts that gravitational waves contain
only two, transverse, polarizations (plus and cross);
observations can constrain or measure the presence of 
longitudinal, scalar, or other polarizations.
In quantum language, this constrains or measures the spin of the graviton. 
This can allow one to test specifically for 
scalar-tensor and other non-tensor theories. 
Possible parity-violating effects~\cite{YunesPV}
could result in differences in the propagation of waves
of different polarizations (right- or left-handed combinations of 
plus and cross).
A more general parameterization of beyond-GR effects in gravitational wave
phenomenology has been developed~\cite{Yunes2009};
see the talk at this conference by Jonathan Gair.
Note that we rely on waveform predictions to identify and detect the weak signals 
in the first place, so a significant deviation from GR predictions
will make it more difficult to detect and thus test for those deviations
(the ``chicken and egg problem'').

\section{Multi-messenger astronomy}

\ssec{Complementary information:}
Energetic astrophysical systems involving ordinary or nuclear matter will emit
both gravitational and electromagnetic radiation over a wide range of 
wavelengths, all carrying different and valuable information about the 
source.  
Gravitational waves tell us about the bulk motion dynamics such as
binary parameters. The GW energetics are a direct probe of the central engine powering 
the energy release. And they can be used as standard sirens to measure luminosity distance
for cosmology.
The spectrum and multi-wavelength light curve of the electromagnetic radiation
gives us the precise sky location, so that we can identify the host galaxy
and galaxy's redshift. The EM afterglow energetics allow us to 
probe the nature of the progenitor star and its gas environment.
Coincident information from GW and EM observations thus provide a much
more complete picture of the progenitor astrophysics.
Combining these observations will also greatly increase the confidence
in the detection above the detector background.

\ssec{The flow of information -- EM triggers:}
EM events can trigger deeper (offline) searches of the GW detector data.
GRBs or SGR flares detected from very wide-field 
space-based X-ray and gamma ray telescopes give a 
precise time and sky location of the event;
this permits a targeted search for coincident GW signals
(in data ``in the can''), reducing the noise contamination in GW detector network
with respect to all-sky, all-time searches.
detection thresholds can be set lower, allowing for deeper searches. 

\ssec{The flow of information -- GW triggers:}
The GW detector network is sensitive to signals from (more-or-less) the whole sky,
at all times (modulo the detectors' operational duty cycle).
By contrast, ground- and space-based optical and radio telescopes
need to know where to point. 
Low latency GRB alerts have allowed for the discovery of optical afterglows,
greatly increasing our information about the sources.
Similarly, low latency GW alerts can serve a similar function
in the cases where there is no prompt GRB, either because it is not 
beamed towards us or is absent from the source. GW emission cannot be beamed
strongly, as GRBs are believed to be.
But to catch the prompt EM afterglow, GW triggers must be generated 
and broadcast, and telescopes must be pointed, quickly. 
Afterglows can fade on the timescale of hours, or even minutes.
Because GW sky localization is poor, even very wide-field (of order 10 square degree)
optical telescopes may need to take many images (over many days) to observe
the sky region consistent with a GW signal, in the hope of finding
a possible EM counterpart.
This  requires development of low-latency GW detection and sky localization pipelines, 
protocols to pass info, telescope scanning strategies and, ideally, 
coordination of EM observations amongst different telescopes.

\ssec{The flow of information -- GW detections and all-sky telescopes:}
GW detectors see the whole sky and can store all the data indefinitely.
So can neutrino detectors like ANTARES and Ice-Cube; present
and future optical transient survey telescopes like PTF, Pan-STARRS and LSST;
and present and future large-scale optical and radio transient surveys (see 
the talk at this conference by Brennan Hughey on radio transients).
Searches for multi-messenger events near-coincident in time and space
can be done offline, through ``data mining''.

\ssec{Enabling the flow:}
Prototypes for all of these paths have been developed;
see talks at this conference by Ruslan Vaulin, Laura Nuttall, 
Darren White, Brennan Hughey, Imre Bartos, Peter Shawhan, and Sharon Rapaport.
But they need to be fully tested, flawless and ready 
when the advanced detector network comes online in the next few years.
Online, low latency GW search pipelines will identify potential
GW detection candidates within $\sim$1 minute of data taking.
They will estimate and broadcast their time, 
sky location with a sky probability map, false alarm rate, 
and other relevant info like binary masses, frequency band, etc.
Information about confirmed detections will take longer (maybe months 
at first), and will include all of the above, 
plus snippets of the strain data h(t), noise spectral densities of the
detectors at the time of detection, improved parameter estimation, etc.
These will be assembled into a catalog of transient detections
to enable queries of the population distribution in terms of 
any and all measured parameters.

Besides for binary mergers, astrophysical sources of high energy photons
that may also have associated GW bursts include gamma ray bursts (GRBs), 
core-collapse supernovae, magnetars and pulsar glitches.

\ssec{Short-hard GRBs:}
Joint GW and EM observations of short hard GRBs detected by 
space-based gamma ray telescopes can
confirm (or rule out) the merger progenitor model;
permit the study of the progenitor systems, including orientation and GRB beaming;
relate the GW and EM energy release to test detailed models;
and relate merger parameters to hosts (metallicity, star formation rate, etc.)
to test population synthesis models.
As mentioned above, such events can be used to make an
independent, self-calibrating measurement of the Hubble constant at low redshift
(since high redshift GWs from mergers will not be detectable by the coming 
advanced detector network).
Coincident detection of high energy neutrinos produced in the internal shock
of the GRB progenitor explosion will be a spectacular discovery.

\ssec{Long-soft GRBs:}
Joint GW and EM observations of long GRBs can test progenitor models
such as the Collapsar model (core collapse of a massive, 
rapidly spinning and highly magnetized star).
The GRB is known to be strongly beamed. The GW burst will not be
strongly beamed, but may be too weak
to be detected at large distances (10 Mpc or more), and the waveforms
are poorly modeled. If the source is sufficiently close,
the copious low energy neutrinos may be detectable.
See the talk at this conference by Michal Was on GRB-triggered GW burst
searches with the initial detector network.

\ssec{Core-collapse supernovae:}
Typical core-collapse supernovae will be observable in GW bursts
for progenitor stars in our Galaxy; extra-galactic supernovae
will produce detectable GWs only in the most extreme models~\cite{CCSNe}.
Unfortunately, the rate of core-collapse supernovae in our Galaxy
is small, of order 1 in 50 years.
The GW burst is from the core collapse itself, while the supernova light
(detectable from even very distant galaxies, but possibly obscured 
by dust for supernovae in our Galaxy)
appears only when the shock breaks out to the surface of the 
progenitor star, hours later. The observation of the GW burst
provides extremely valuable information about the details
of the core collapse and subsequent shock as it propagates through 
the outer stellar material.
The huge energy release of low-energy neutrinos 
(detectable from supernovae in our local group of galaxies)
provide more valuable information
about how the shock propagates and breaks out.

\ssec{Magnetars:}
Highly magnetized neutron stars ($B \sim 10^{15}\, G$)
are observed as Soft Gamma Repeaters
(SGRs)  and Anomalous X-ray pulsars (AXPs). 
On the order of 20 are known in our Galaxy.
They emit short bursts of X-rays and soft gamma-rays ($\sim 10^{42}$ erg), 
sometimes giant flares ($10^{44}-10^{46}$ erg) or storms of bursts, 
and QPOs in the giant flare tails ($10 - 2000$ Hz).
The energy release may excite neutron star crustal 
f-modes in the kHz region, ringing down by GW emission, but with unknown energy.
Torsional (shear) s-mode ``core-quakes'' may release significantly more GW energy.
Searches with initial detector data ~\cite{magnetarsearches}
yielded limits on GW energy emission 
significantly higher than EM emission 
($10^{48}-10^{50}$ erg);
see the talk at this conference by Ray Frey.
With advanced detectors that will be 10 times more sensitive,
we can expect roughly a factor 100 better sensitivity in terms of emitted energy;
and we can hope to get lucky with a nearby magnetar burst.

\ssec{Pulsar Glitches:}
Pulsar glitches may be caused by star-quakes, 
or the sudden transfer of angular momentum from a superfluid core to a solid crust.
These energetic events may excite quasi-normal mode oscillations 
in the neutron star interior, coupling to GW emission.
An Initial LIGO search for GW ringdown signals (1--3 kHz) 
from a Vela pulsar glitch in 2006~\cite{Velaglitch} 
yielded upper limits in the range of $10^{44}-10^{46}$ erg;
see the talk at this conference by James Clark.
Many more pulsars will be discovered and monitored in the coming years 
with LOFAR and  SKA. Nearby, glitchy ones will be promising
candidates for coincident GW emission detectable by the advanced detector network.

\section{Summary and conclusions}

Gravitational wave transients from extremely energetic astrophysical events
are a rich trove of information about the progenitor system,
both on their own and in coincidence with information
from gamma ray, X-ray, optical, radio, and/or neutrino emission.

We look forward to the coming advanced detector era:
the discovery and exploration of the GW sky, 
joint observations and discoveries with EM and neutrino telescopes, 
and the emergence of a rich new branch of astrophysics.

We can anticipate much of what is to come, but we will learn the most
from the discoveries that are as yet unexpected.

\section{Acknowledgments}

The author thanks the conference organizers for an extremely
stimulating and enjoyable meeting.
The author is representing the LIGO Scientific Collaboration
and the Virgo Collaboration, and gratefully acknowledges the rich trove
of information produced by his colleagues that has enabled this work.

The authors gratefully acknowledge the support of the United States
National Science Foundation for the construction and operation of the
LIGO Laboratory, the Science and Technology Facilities Council of the
United Kingdom, the Max-Planck-Society, and the State of
Niedersachsen/Germany for support of the construction and operation of
the GEO600 detector, and the Italian Istituto Nazionale di Fisica
Nucleare and the French Centre National de la Recherche Scientifique
for the construction and operation of the Virgo detector. 
The authors
also gratefully acknowledge the support of the research by these
agencies and by the Australian Research Council, 
the International Science Linkages program of the Commonwealth of Australia,
the Council of Scientific and Industrial Research of India, 
the Istituto Nazionale di Fisica Nucleare of Italy, 
the Spanish Ministerio de Educaci\'on y Ciencia, 
the Conselleria d'Economia Hisenda i Innovaci\'o of the
Govern de les Illes Balears, the Foundation for Fundamental Research
on Matter supported by the Netherlands Organisation for Scientific Research, 
the Polish Ministry of Science and Higher Education, the FOCUS
Programme of Foundation for Polish Science,
the Royal Society, the Scottish Funding Council, the
Scottish Universities Physics Alliance, The National Aeronautics and
Space Administration, the Carnegie Trust, the Leverhulme Trust, the
David and Lucile Packard Foundation, the Research Corporation, and
the Alfred P. Sloan Foundation.
This paper has LIGO document number LIGO-P1100189.

\end{document}